# THE STATISTICAL SIGNIFICANCE OF MULTIVARIATE HAWKES PROCESSES FITTED TO LIMIT ORDER BOOK DATA

ROGER MARTINS
DIETER HENDRICKS*

*School of Computer Science and Applied Mathematics, University of the Witwatersrand
Johannesburg, WITS 2050, South Africa*

Hawkes processes have seen a number of applications in finance, due to their ability to capture event clustering behaviour typically observed in financial systems. Given a calibrated Hawkes process, of concern is the statistical fit to empirical data, particularly for the accurate quantification of self- and mutual-excitation effects. We investigate the application of a multivariate Hawkes process with a sum-of-exponentials kernel and piecewise-linear exogeneity factors, fitted to liquidity demand and replenishment events extracted from limit order book data. We consider one-, two- and three-exponential kernels, applying various tests to ascertain goodness-of-fit and stationarity of residuals, as well as stability of the calibration procedure. In line with prior research, it is found that performance across all tests improves as the number of exponentials is increased, with a sum-of-three-exponentials yielding the best fit to the given set of coupled point processes.

## 1. Introduction

Hawkes (1971) introduced a class of multivariate point processes with a stochastic intensity vector, incorporating event-occurrence clustering behaviour in a coupled system. Initial applications used calibrated Hawkes processes to measure the conditional intensities of earthquakes and aftershocks, based on recorded data (Vere-Jones (1970), Ogata (1988, 1999)). In financial markets, empirical studies of market microstructure have highlighted apparent clustering of limit order book events at tick scale, with some event intensities exhibiting dependent behaviour (Biais et al. (1995), Gould et al. (2013), Abergel et al. (2015)).

The simplest Hawkes processes are univariate, considering a single event type and its temporal dependence on prior events. These models contain an *exogenous* or *baseline intensity* component, which corresponds to the intensity of events that is not influenced by the occurrence of prior events, and an *endogenous* intensity component, which corresponds to the increased intensity of *child* events that occurs

*Corresponding author: dieter.hendricks@students.wits.ac.za

2  *R. Martins, D. Hendricks*

after an exogenous *parent* event, thus capturing the clustering of events that one often expects in financial data. A multivariate Hawkes process, then, extends this to include multiple event types, where we further define *self-excitation* of one event type influencing more of its own type of event, and *mutual-* or *cross-excitation* of one event precipitating the occurrence of other types of events.

**Definition 1.1. Multivariate Hawkes process**
Consider a point process $N(t)$ such that

$$\mathbb{P}[\triangle N(t) = 1 | N(s)_{s \leq t}] = \lambda(t)\triangle_t + o(\triangle_t) \text{ and}$$
$$\mathbb{P}[\triangle N(t) > 1 | N(s)_{s \leq t}] = o(\triangle_t).$$

For a multivariate point process $N(t) = \{N_r(t) : r = 1, ..., R\}$, the $r^{th}$ intensity function of the $R$-variate mutually-exciting Hawkes process is given by

$$\lambda_r(t) = \mu_r(t) + \int_{-\infty}^{t} \sum_{r=1}^{R} \phi_{r,i}(t-u) dN_i(u)$$

where

$\mu_r(t)$ is the time-dependent baseline intensity for the $r^{th}$ event type

$\phi_{r,i}(t)$ is the kernel function, which encodes the dependency on

prior events of type $i$ and satisfies the following conditions (Bacry et al. (2015)):

1) *Component-wise positive, i.e. $\phi_{r,i}(t) \geq 0$ for each $1 \leq r, i \leq R$*

2) *Component-wise causal, i.e. if $t < 0$, then $\phi_{r,i}(t) = 0$ for each $1 \leq r, i \leq R$*

3) *Each component belongs to the space of $L^1$-integrable functions.*

The point process $N(t)$ and intensity vector $\lambda(t)$ together characterise the Hawkes process.

Bacry et al. (2015) provide a comprehensive review article highlighting the many applications of Hawkes processes in finance. Bowsher (2005) considered one of the first applications, where a bivariate point process of the timing of a stock's trade price and mid-quote changes was used to model volatility clustering on the New York Stock Exchange. A key phenomenon investigated using Hawkes processes is *endogeneity* in financial markets (Filimonov and Sornette (2012, 2013), Hardiman et al. (2013), Hardiman and Bouchaud (2014)). Empirical observation reveals that, in certain instances, market prices change too quickly to be strictly attributed to the flow of pertinent information, and thus evade explanation in classic economic theory (Bowsher (2005)). By considering the ratio of exogenous parent events to endogenous events, it is possible to obtain a measure of market reflexivity (Filimonov and Sornette (2012)).

Degryse et al. (2005), Large (2007), Bacry and Muzy (2014a) used a multivariate Hawkes process to quantify the *resiliency* of a limit order book (LOB), viz. the propensity for quote replenishment following a liquidity demand event. By characterising and extracting key liquidity demand and replenishment events from a limit order book, and using an appropriate choice of kernel to encode temporal dependence of events, Large (2007) claims it is possible to use a calibrated Hawkes process to calculate the probability and expected half-life of quote replenishment following a liquidity demand event.

We identify key aggressive liquidity demand and replenishment events, enumerating empirical event point processes for model calibration and quantification of LOB resiliency. The primary investigation in this paper is to find a kernel for the multivariate Hawkes process which provides a significant fit to the empirical data extracted from the Johannesburg Stock Exchange (JSE). We note that this work is very much in the same spirit as Lallouache and Challet (2016), however we are determining the statistical significance of *multivariate* Hawkes processes fitted to LOB event data.

This paper proceeds as follows: Section 2 describes the key liquidity demand and replenishment point processes extracted from the limit order book data, which were used for calibrating the Hawkes processes. Section 3 highlights some candidate kernels which can be used to encode temporal dependence amongst system events. Section 4 discusses the derivation of the likelihood function when using the sum-of-expenentials kernel, used in this analysis. Section 5 discusses the model calibration procedure and implementation. Section 6 discusses the distribution and stationarity tests for the residuals to assess each model's goodness-of-fit to the provided data. Section 7 discusses the data and results for this analysis and Section 8 provides some concluding remarks.

## 2. Enumerating empirical event point processes using tick data

Typical limit order book (LOB) events include trades, new quotes, quote modifications and quote cancellations (Abergel et al. (2015)). Following the suggestions by Large (2007) and Biais et al. (1995) and taking into account the nature of our data, we define 4 key *aggressive* liquidity demand and replenishment event types which will be used to characterise order book resiliency:

- **Type 1:** *A buy trade that moves the offer*
  The first of the two liquidity demand events, we define an aggressive buy trade as one where the trade price is greater than the best offer price, or it is equal to the best offer, but the volume is greater than that available at the current best offer. Such trades are considered aggressive since they materially alter the shape of the limit order book, pushing the best offer price higher, widening the spread and removing liquidity. Formally, with $P$ representing the trade price, $A$ the prevailing best offer, and $V_P$, $V_A$ the respective volumes, we express this event type as the following subset of

limit order book events:
$$(P > A) \cup \big[(P = A) \cap (V_P \geq V_A)\big]$$

- **Type 2:** *A sell trade that moves the bid*
  The second of the two liquidity demand events is the aggressive sell trade, where the trade price is lower than the current best bid price, or it is equal to the best bid price, but the volume of the trade is greater than that available at the best bid. Similarly, this trade is considered aggressive since it alters the structure of the limit order book by pushing the best bid down, widening the spread and removing liquidity. Referring to the aforementioned notation, only adding that $B$ refers to the best available bid price, we formally characterise aggressive sell trades as the following subset:
  $$(P < B) \cup \big[(P = B) \cap (V_P \geq V_B)\big]$$

- **Type 3:** *A bid between the quotes*
  The first of the two resiliency events, the aggressive bid is a bid between the current best bid and offer. It is considered aggressive since it alters the structure of the limit order book, pushing up the best bid, reducing the spread and providing liquidity through more competitive prices and added volume. With $B_*$ being the incoming bid quote, and $B_p$ the prevailing best bid, we formally characterise the aggressive bid using similar notation to before, as the following subset:
  $$(B_* > B_p)$$

- **Type 4:** *An offer between the quotes*
  The second of the two resiliency events, an aggressive offer is defined as an offer quote between the current best bid and offer, considered aggressive since it narrows the gap between the prevailing best bid and offer, providing liquidity. With $A_*$ being the incoming offer quote, and $A_p$ the prevailing best offer, we formally characterise the aggressive offer quotes as the following subset:
  $$(A_* < A_p)$$

While not critical for our study of LOB resiliency, we also identify and classify the following *passive* event types for completeness:

- **Type 5:** *Passive buy trade*
  These are buy trades which do not negatively affect LOB liquidity, as they do not affect the prevailing spread. If $P$ is the trade price, $A$ the prevailing offer quote, and $V_P$ and $V_A$ their respective volumes, passive buy trades will be classified as:
  $$\big[(P = A) \cap (V_P < V_A)\big]$$

- **Type 6:** *Passive sell trade*
  These are sell trades which do not negatively affect LOB liquidity, as they do not affect the prevailing spread. If $P$ is the trade price, $B$ the prevailing bid quote, and $V_P$ and $V_B$ their respective volumes, passive sell trades will be classified as:
  $$\big[(P = B) \cap (V_P < V_B)\big]$$

- **Type 7:** *Passive bid quote*
  These are bid quotes which enter the LOB at a level higher than level-1, and hence do not affect the prevailing spread. If $B_*$ is the incoming bid quote and $B_p$ is the prevailing bid quote, passive bid quotes will be classified as:
  $$(B_* < B_p)$$

- **Type 8:** *Passive sell quote*
  These are offer quotes which enter the LOB at a level higher than level-1, and hence do not affect the prevailing spread. If $A_*$ is the incoming offer quote and $A_p$ is the prevailing offer quote, passive offer quotes will be classified as:
  $$(A_* > A_p)$$

Figures 1, 2 and 3 illustrate the measured empirical intensities for the 4 key event types, demonstrating event clustering and mutual-excitation over typical morning, midday and afternoon periods.

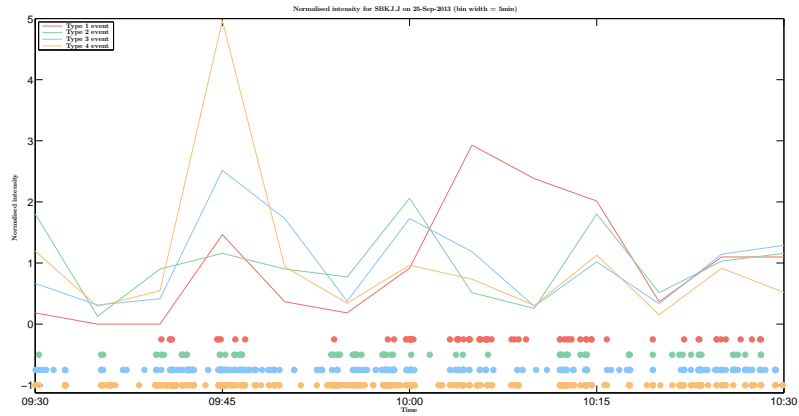

Fig. 1. Empirical event intensities for SBK of each of the 4 key event types over a **morning period**, demonstrating event clustering and mutual-excitation. The dots show the arrival times of the events and the lines show the 5-minute event intensities.

6   *R. Martins, D. Hendricks*

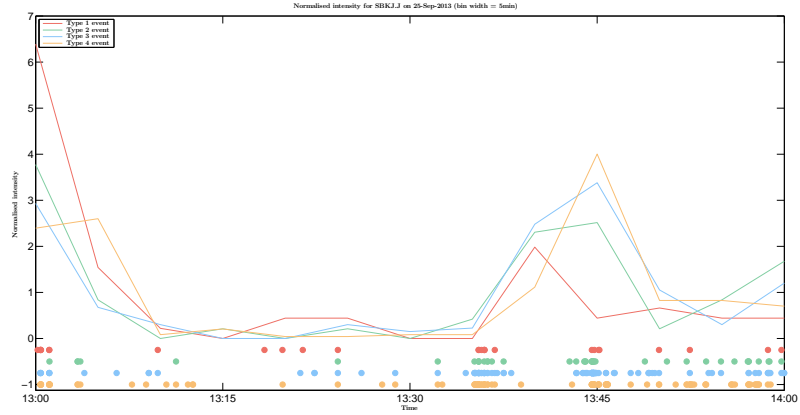

Fig. 2. Empirical event intensities for SBK of each of the 4 key event types over a **midday period**, demonstrating event clustering and mutual-excitation. The dots show the arrival times of the events and the lines show the 5-minute event intensities.

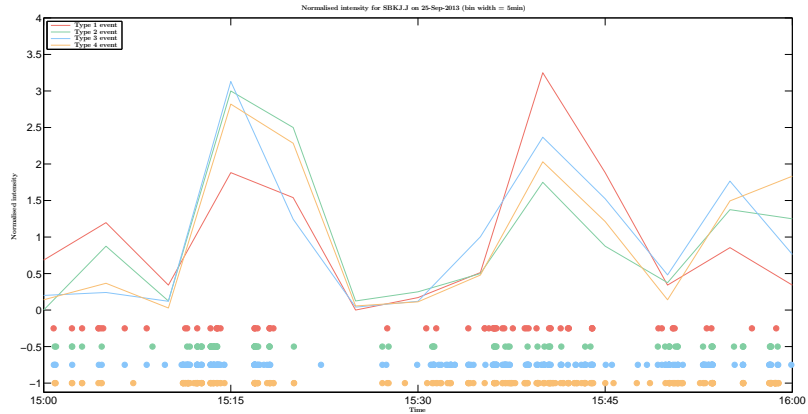

Fig. 3. Empirical event intensities for SBK of each of the 4 key event types over an **afternoon period**, demonstrating event clustering and mutual-excitation. The dots show the arrival times of the events and the lines show the 5-minute event intensities.

## 3. Candidate kernels for encoding temporal dependence

A number of kernels have been proposed to model temporal dependence of events, informed by the application or the dynamics of the data being modelled (Hardiman and Bouchaud (2014)). While, in principle, any kernel satisfying the conditions in Definition 1.1 can be used, four candidate kernels have typically been considered for financial applications (Large (2007), Bacry et al. (2015), Lallouache and Challet (2016)):

*The statistical significance of multivariate Hawkes processes fitted to limit order book data*    7

- *Sum-of-exponentials*:

$$\phi_M(t) = \sum_{i=1}^{M} \alpha_i e^{-t/\tau_i}$$

  where $M$ is the number of exponentials, $\alpha_i$ is the amplitude of the $i^{th}$ kernel and $\tau_i$ is the timescale of the $i^{th}$ kernel. The *branching ratio* is calculated as $n = \sum_{i=1}^{M} \alpha_i \tau_i$.

- *Approximate power law*:

$$\phi_M(t) = \frac{n}{Z} \sum_{i=1}^{M-1} a_i^{-(1+\epsilon)} e^{-\frac{t}{a_i}}$$

  where $a_i = \tau_0 m^i$, $M$ is the range of approximation and $m$ its precision. $Z$ is defined such that $\int_0^\infty \phi_M(t) dt = n$, $n$ is the *branching ratio*, $\epsilon$ is the tail exponent and $\tau_0$ is the smallest timescale.

- *Approximate power law with short lag cut-off* (Hardiman et al. (2013)):

$$\phi_M(t) = \frac{n}{Z} \Big( \sum_{i=1}^{M-1} a_i^{-(1+\epsilon)} e^{-\frac{t}{a_i}} - S e^{-\frac{t}{a_{-1}}} \Big)$$

  where the definition is the same as the *approximate power law*, with the addition of a smooth exponential drop for lags shorted than $\tau_0$. $S$ is defined such that $\phi_M(0) = 0$.

- *Lallouache-Challet power law and exponential* (Lallouache and Challet (2016)):

$$\phi_M(t) = \frac{n}{Z} \Big( \sum_{i=1}^{M-1} a_i^{-(1+\epsilon)} e^{-\frac{t}{a_i}} + b e^{-\frac{t}{\tau}} \Big)$$

  which is the *approximate power law* with an additional exponential term with free parameters $b$ and $\tau$. This permits greater freedom in the structure of time scales.

We will consider the *sum-of-exponentials* kernel, as this will allow us to quantify limit order book (LOB) resiliency by measuring the branching ratio for particular liquidity demand/replenishment event pairs, as well as the expected half-life of replenishment. Large (2007) made use of a 2-exponential kernel to quantify LOB resiliency, however it is unclear whether two exponentials is appropriate for our events dataset. We will thus investigate the *goodness-of-fit* for $M = 1, 2$ and 3 to determine the appropriate form of the kernel, given the event processes enumerated from our dataset. We note that recent studies have considered efficient non-parametric calibration procedures for estimating the branching ratio of a symmetric Hawkes process (Bacry et al. (2012), Bacry and Muzy (2014b)). While we have taken measures to promote stability in our calibration procedure (discussed in Section 5),

8   *R. Martins, D. Hendricks*

further studies should consider the application of these non-parametric estimators and their ability to capture results of the *full* model when applied to data extracted from the JSE LOB.

## 4. Deriving a maximum likelihood estimator with the sum-of-exponentials kernel

We will consider the exposition shown in Lallouache and Challet (2016) and extend it to the multivariate case. The sum-of-exponentials kernel is defined as

$$\phi_M(t) = \sum_{i=1}^{M} \alpha_i e^{-t/\tau_i}.$$

The $M$ term refers to the number of exponentials to be summed for each event type, $\alpha$ is the individual exponential's unscaled intensity, $\tau$ refers to the particular timescale associated to the intensity of one exponential, in contrast to the commonly presented $\beta$ that refers to the multiplicative inverse of $\tau$, being the decay. Importantly, we note that the *branching ratio*, $n$, of a particular event type is expressed as

$$n = \sum_{i=1}^{M} \alpha_i \tau_i. \tag{4.1}$$

This branching ratio corresponds to the number of children events a parent event is expected to have. A branching ratio greater than one (*super-critical*) will quickly explode with events, a branching ratio equal to one (*critical*) is a special case where a family will live indefinitely without exploding, as long as $\mu = 0$, and a branching ratio less than one (*sub-critical*) refers to a process where each family of clustered events will eventually die out.

To calculate the half-life of a given intensity effect $\alpha_i$, we solve

$$\frac{1}{2} = e^{-\frac{t^i_{\frac{1}{2}}}{\tau_i}} \implies t^i_{\frac{1}{2}} = \tau_i \ln(2), \tag{4.2}$$

with the total half-life given by

$$t_{\frac{1}{2}} = \sum_{i=1}^{M} \tau_i \ln(2). \tag{4.3}$$

The sum-of-exponentials kernel thus yields the following intensity function:

$$\lambda_M(t) = \mu + \int_0^t \sum_{i=1}^{M} \alpha_i e^{-\frac{t-u}{\tau_i}} dN(u).$$

In particular, we are considering the four-variate case, since we are interested in key aggressive liquidity demand and replenishment events to quantify resiliency

*The statistical significance of multivariate Hawkes processes fitted to limit order book data*   9

(see Section 2 below). With $r = 1, 2, 3, 4$ referring to the event type of interest $\dot{r}$, and assuming a time-dependent baseline intensity, $\mu$, we have

$$\lambda_{\dot{r},M}(t) = \mu_{\dot{r}}(t) + \int_0^t \sum_{r=1}^4 \sum_{i=1}^M \alpha_{r,i} e^{-\frac{t-u}{\tau_{r,i}}} dN_{\dot{r}}(u). \tag{4.4}$$

We refer to the known log-likelihood function for Hawkes processes with exponential or power-law kernels Ozaki (1979):

$$\ln \mathcal{L}(t_1, ..., t_n | \theta) = -\int_0^T \lambda(t|\theta)dt + \int_0^T \ln \lambda(t|\theta)dN(t)$$

noting that

$$\int_0^t h(s)dN(s) = \sum_{t_i < t} h(t_i).$$

Thus we derive the log-likelihood as

$$\ln \mathcal{L}_{\dot{r}}(\theta) = -\int_0^T \mu_{\dot{r}}(t)dt - \sum_{r=1}^4 \sum_{i=1}^M \alpha_{r,i}\tau_{r,i} \sum_{t_j < T} \left(1 - e^{-\frac{T-t_j}{\tau_{r,i}}}\right)$$
$$+ \sum_{t_j < T} \ln \left[\mu_{\dot{r}}(t_j) + \sum_{r=1}^4 \sum_{i=1}^M \alpha_{r,i} \sum_{t_{j'} < t_j} e^{-\frac{t_j - t_{j'}}{\tau_{r,i}}}\right].$$

Using the following recursive relationships Ozaki (1979), assuming that $\dot{r} = 1$:

$$R_{1,i}(j) = e^{-\frac{t_j - t_{j-1}}{\tau_{1,i}}} \left(1 + R_{1,i}(j-1)\right)$$

and, for mutual event types $r = 2, 3, 4$, letting $\tilde{k} = \sup[k'|t_{k'} < t_j]$:

$$R_{r,i}(j) = e^{-\frac{t_j - t_{j-1}}{\tau_{r,i}}} R_{r,i}(j-1) + \sum_{[k'|t_{j-1} \leq t_{k'} < t_j]} e^{-\frac{t_j - t_{k'}}{\tau_{r,i}}}$$

Thus, substituting back into the log-likelihood function, we obtain

$$\ln \mathcal{L}_{\dot{r}}(\theta) = -\int_0^T \mu_{\dot{r}}(t)dt - \sum_{r=1}^4 \sum_{i=1}^M \alpha_{r,i}\tau_{r,i} \sum_{t_j < T} \left(1 - e^{-\frac{T-t_j}{\tau_{r,i}}}\right)$$
$$+ \sum_{t_j < T} \ln \left[\mu_{\dot{r}}(t_j) + \sum_{r=1}^4 \sum_{i=1}^M \alpha_{r,i} R_{r,i}(j)\right] \tag{4.5}$$

The log-likelihood function in Equation 4.5 will be implemented for parameter calibration.

## 5. Calibration of model parameters

To quantify LOB resiliency, we require the calibration of the $\mu$, $\alpha$ and $\tau$ parameters in Equation 4.5. We used MATLAB to develop a non-linear constrained optimisation routine to find the parameters which maximise the likelihood function specified in Equation 4.5. In particular, we used a sequential quadratic programming algorithm to iteratively adjust candidate parameter values until a best approximation to the maximum likelihood estimator is found, within a given tolerance. To promote finding a global solution and stability in algorithm results, we use a genetic algorithm with Equation 4.5 as the objective function to find feasible parameter values to initialise the optimisation routine. This allows us to narrow the search space, before using the optimisation to refine the calibration.

While the univariate Hawkes process can exploit the recursive relationship for the log-likelihood calculation, reducing the computational complexity from $\mathcal{O}(N^2)$ to $\mathcal{O}(N)$ (Ozaki (1979), Lallouache and Challet (2016)), this advantage does not translate to the multivariate case. We made use of a number of vectorisation enhancements to improve the computational efficiency of the implemented algorithm, as shown in Figure 4. These indicate that our implementation scales well as a function of both number of events and number of exponentials in the chosen kernel, compared to a naïve for-loop implementation. The details of the full implementation can be found in a study by Martins (2015).

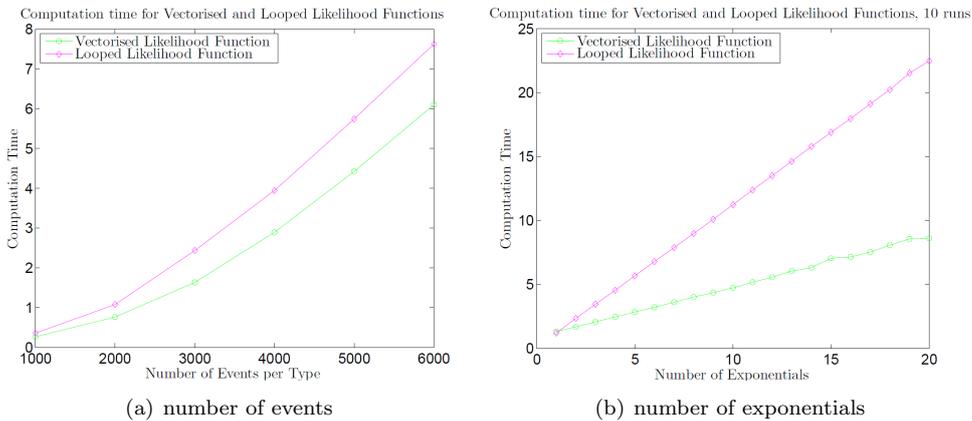

(a) number of events    (b) number of exponentials

Fig. 4. **Computation time (in minutes): vectorisation vs for-loop**. Shows average computation time as a function of the number of each event type, as well as number of exponentials in the kernel.

Once the parameters have been calibrated to the events data, we can obtain values for the branching ratio (Equation 4.1) and half-life (Equation 4.3), and the particular expression of Equation 4.4 that defines the intensity of the Hawkes process, which can in turn be used to calculate the residuals for testing.

Before any inferences were made from calibrated parameter values, we verified the accuracy of our implemented calibration scheme. To do this, we performed extensive simulations of multivariate Hawkes processes with known, induced parameters using the *intensity-based approach* promoted by Lewis and Shedler (1979), Dassios and Zhao (2013), generating a set of events data. The calibration scheme was then used to recover the induced parameters. The successful recovery of induced parameters from simulated data verified the calibration scheme. Details of this can be found in a study by Mazibuko (2014).

## 6. On the choice of $M$ (number of exponentials)

In order to determine the appropriate number of exponentials in our kernel, we performed a number of *goodness-of-fit* tests for $M = 1, 2$ and $3$ following calibrations to our dataset of events. We utilise the result in Theorem 6.1 to convert our calibrated Hawkes processes into time-deformed *compensator* functions, viz. sequences of residuals which are exponentially distributed with mean 1.

**Theorem 6.1 (Multivariate random time change).** *Consider the R sequences of generalised residuals $\{e_i^r(\theta)\}$, $r = 1, ..., R$, where*

$$e_i^r(\theta) := \int_{t_i^{(r)}}^{t_{i+1}^{(r)}} \lambda_r(s, \theta) ds, \qquad (6.1)$$

*the integrand is the intensity for type $r$ events, and $(t_i^{(r)}, t_{i+1}^{(r)}]$ is interval between adjacent type $r$ events.*
*When $\theta$ is the set of true parameters, each sequence $e_i^r(\theta)$ is an independently distributed exponential random variable with mean 1.*

**Proof.** See Bowsher (2005) for a detailed discussion and proof. □

We can then use statistical tests to determine whether the residuals are in fact independent and exponentially distributed with mean 1, where the kernel offering the *best fit* will be used in the analyses that follow. We used four candidate tests to assess the distribution and stationarity of the residuals (Large (2007), Hardiman et al. (2013), Lallouache and Challet (2016)):

(1) *Kolmogorov-Smirnov (KS) test* (Kolmogorov (1933), Smirnov (1948)): This test compares the empirical distribution of residuals to cumulative distribution of an exponential with mean 1. An *acceptance* of the null hypothesis $H_0$ indicates empirical residuals are exponentially distributed at the specified significance level.
(2) *Excess Dispersion (ED) test* (Engle and Russell (1998)): This test confirms whether the residuals have unit variance, as would be expected if they followed an exponential with mean 1. An *acceptance* of the null hypothesis $H_0$ indicates the variance of empirical residuals is 1 at the specified significance level.

12  *R. Martins, D. Hendricks*

(3) *Ljung-Box Q (LBQ) test* (Ljung and Box (1978)): This test for stationarity confirms independence of increments, where again, *acceptance* of the null hypothesis $H_0$ indicates the residuals are stationary.
(4) *Kwiatkowski-Phillips-Schmidt-Shin (KPSS) test* (Kwiatowski et al. (1992)): This is a secondary test for stationarity for confirmation of the LBQ results, where *acceptance* of the null hypothesis $H_0$ indicates the residuals are stationary.

The KS, ED and LBQ statistical tests were performed at a 1% significance level, while the KPSS test was performed at the 5% significance level.

## 7. Data and Results

### 7.1. *Data*

The data for our analysis consisted of transactions and market depth quotes for two stocks listed on the Johannesburg Stock Exchange (JSE), sourced from Thomson Reuters Tick History (TRTH). The methodology described in Section 2 was used to extract event point processes for model calibration.

**Standard Bank (SBK)** Stability tests were performed on Standard Bank tick data from 05 March 2012 to 09 March 2012. For the four events of interest, this data set contained a total of 13213 events over all five days. Additional statistics pertaining to the data set are presented in the following table, recalling that all event types are aggressive:

| Event Type | Total in Set | Mean per Day | SD per Day |
|---|---|---|---|
| Buy (*Type 1*) | 1583 | 316.6 | 35.3808 |
| Sell (*Type 2*) | 1511 | 302.2 | 53.7513 |
| Bid (*Type 3*) | 5173 | 1034.6 | 226.9324 |
| Ask (*Type 4*) | 4946 | 989.2 | 368.6227 |

Table 1. **Standard Bank, 05 March 2012 to 09 March 2012**

**Growthpoint Properties Limited (GRT)** Daily calibrations were performed on two sets of Growthpoint data; one from 01 September 2013 to 27 September 2013 with a total of 39347 events over 19 days, and the other from 30 September 2013 to 31 October 2013 with 31090 events over 24 days.

*The statistical significance of multivariate Hawkes processes fitted to limit order book data*   13

| Event Type | Total in Set | Mean per Day | SD per Day |
|---|---|---|---|
| Buy (*Type 1*) | 5332 | 280.6316 | 102.3139 |
| Sell (*Type 2*) | 4962 | 261.1579 | 115.2380 |
| Bid (*Type 3*) | 13237 | 696.6842 | 248.6008 |
| Ask (*Type 4*) | 15816 | 832.4211 | 483.2089 |

Table 2. **Growthpoint Properties Limited, 01 September 2013 to 27 September 2013**

| Event Type | Total in Set | Mean per Day | SD per Day |
|---|---|---|---|
| Buy (*Type 1*) | 4019 | 167.4583 | 65.5054 |
| Sell (*Type 2*) | 4489 | 187.0417 | 56.3417 |
| Bid (*Type 3*) | 10431 | 434.6250 | 140.3644 |
| Ask (*Type 4*) | 12151 | 506.2917 | 200.0264 |

Table 3. **Growthpoint Properties Limited, 30 September 2013 to 31 October 2013**

### 7.2. *Results*

Given the onerous calibration procedure and high number of model parameters, we first performed a *stability test* to assess the consistency of our implementation. This was performed on a relatively small dataset of SBK point processes (Table 1), where the calibration and subsequent statistical tests were performed 100 times on the same data. Following confirmation of the stability of the calibration procedure, we then moved on to the *goodness-of-fit tests*, where we assess the models under the 3 candidate kernels and examine the residuals when applied to the GRT point processes (Tables 2 and 3). Here, each day is treated as an independent realisation of the multivariate point process, hence the models are calibrated each day with summary results shown. We separate the analysis into two calendar month periods (01 September 2013 to 27 September 2013 and 30 September 2013 to 31 October 2013), as there was a significant change in the fee structure on the JSE on 30 September 2013 (Harvey et al. (2016)), which may have affected prevailing LOB dynamics.

#### 7.2.1. *Stability tests*

Table 4 and Figure 5 summarise the results of the stability tests. They suggest positive test performance across the three kernels, with the exception of the single exponential's stationarity tests, where both the LBQ and KPSS tests failed more often compared to the other two kernels. We observe that as the number of exponentials increases, more tests are passed, with the sum of three exponentials passing the majority of its tests.

Curiously, we see that the mean *p*-value for the KS tests decreases with the number of exponentials in this particular set of results, but the test pass rate in-creases. Figure 6 demonstrates that, despite scoring high $p$-values for some events, the single exponential has the lowest $p$-values and most failures in event 1, bringing its overall pass rate down.

| M | KS $H_0$ | KS $p$ | ED $H_0$ | ED $p$ | LBQ $H_0$ | LBQ $p$ | KPSS $H_0$ | KPSS $p$ |
|---|---|---|---|---|---|---|---|---|
| 1 | 0.9750 | 0.1660 | 0.950 | 0.3685 | 0.5200 | 0.4359 | 0.5000 | 0.0678 |
| 2 | 0.9825 | 0.0839 | 0.990 | 0.5571 | 0.8818 | 0.4437 | 0.8900 | 0.0927 |
| 3 | 0.9975 | 0.0671 | 0.995 | 0.6258 | 0.9460 | 0.4538 | 0.9600 | 0.0972 |

Table 4. **Stability test statistics by kernel**. $H_0$ columns present percentage of null hypotheses accepted, $p$ columns present mean $p$-values over all tests for given kernel.

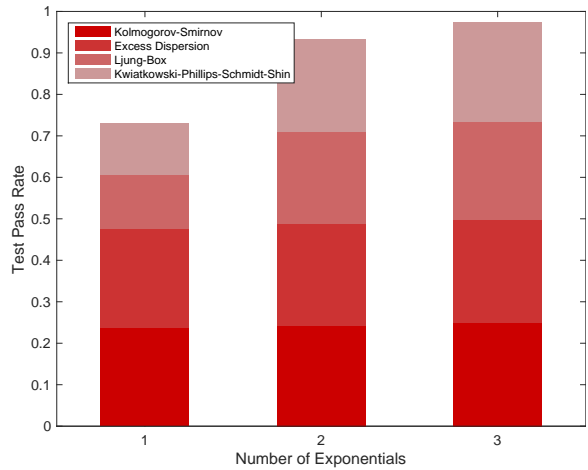

Fig. 5. **Stability test pass rate by kernel** Stacked bars represent the percentage of passes for each statistical test, as achieved by the number of exponentials summed in the kernel of interest. Tests were performed on 100 calibrations of the same SBK dataset.

### 7.2.2. *Goodness-of-fit tests*

Tables 5 and 6 and Figure 7 show the *goodness-of-fit* results for empirical event processes of a candidate stock (GRT) over 2 different periods, for $M = 1, 2$ and 3. We have used to definitions in Section 2 to extract key aggressive liquidity demand and replenishment processes from the LOB data. Each day is considered as an independent realisation of the multivariate point process, thus a separate calibration is performed for the set of events associated with each day in the dataset. This allows us to construct a distribution of null hypothesis *acceptances* and $p$-values for each test, and for each $M$. For example, column "KS $H_0$" indicates the proportion of KS tests which confirmed residuals were exponentially distributed with unit mean, whereas column "KS $p$-value" shows the average $p$-value for these tests. We see from these results that the 3-exponential kernel offers a better fit to empirical data than the 1- and 2-exponential kernel. This is confirmed by the highest number of null hypothesis acceptances and highest $p$-values across all statistical tests. We will thus use the 3-exponential kernel in our study of LOB resiliency.

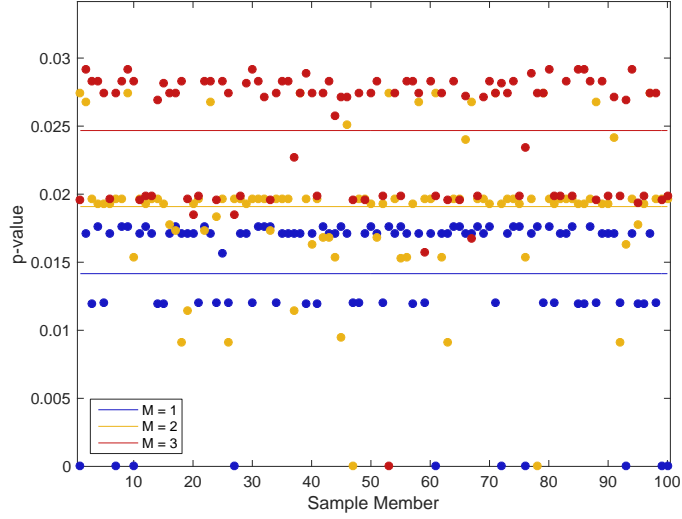

Fig. 6. **Stability test Kolmogorov-Smirnov $p$-values** Comparison of KS test $p$-values obtained by kernels with $M$ exponentials over the 100 calibrations.

| M | KS $H_0$ | KS $p$ | ED $H_0$ | ED $p$ | LBQ $H_0$ | LBQ $p$ | KPSS $H_0$ | KPSS $p$ |
|---|---|---|---|---|---|---|---|---|
| **1** | 0.3421 | 0.0332 | 0.2500 | 0.0628 | 0.7000 | 0.2484 | 0.5658 | 0.0619 |
| **2** | 0.5658 | 0.0978 | 0.4342 | 0.1311 | 0.7184 | 0.2902 | 0.5395 | 0.0589 |
| **3** | 0.6184 | 0.1384 | 0.5789 | 0.1720 | 0.7326 | 0.2945 | 0.5526 | 0.0604 |

Table 5. Daily goodness-of-fit test statistics by kernel, **GRT 01 September 2013 to 27 September 2013**

| M | KS $H_0$ | KS $p$ | ED $H_0$ | ED $p$ | LBQ $H_0$ | LBQ $p$ | KPSS $H_0$ | KPSS $p$ |
|---|---|---|---|---|---|---|---|---|
| **1** | 0.5000 | 0.0884 | 0.3958 | 0.0928 | 0.7475 | 0.2733 | 0.5729 | 0.0627 |
| **2** | 0.6771 | 0.1712 | 0.5938 | 0.1784 | 0.7783 | 0.2790 | 0.6146 | 0.0643 |
| **3** | 0.7917 | 0.2599 | 0.7292 | 0.2658 | 0.8133 | 0.2996 | 0.6250 | 0.0669 |

Table 6. Daily goodness-of-fit test statistics by kernel, **GRT 30 September 2013 to 31 October 2013**

16  *R. Martins, D. Hendricks*

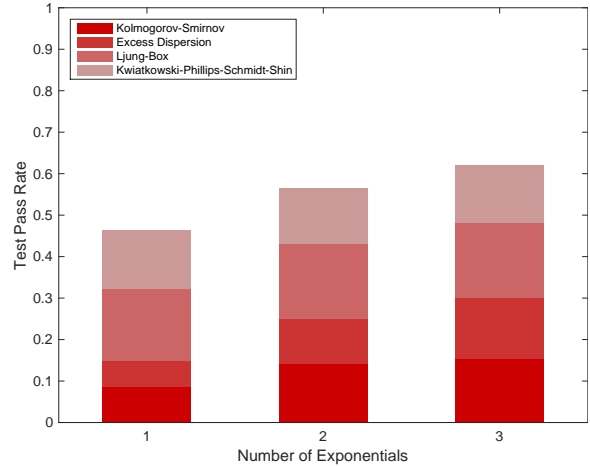

Fig. 7. **Daily Test Pass Rate By Kernel** Tests were performed on calibrations of each day of the GRT dataset. This figure combines the distributions of test pass rates across both periods.

7.2.3. *Discussion on calibrated parameters*

In addition to the statistical testing to establish model accuracy, this analysis includes a discussion of the interpretation of calibrated parameters, and whether or not the calibrated parameters present reasonable values.

**Motivating use of time-dependent baseline intensity** To motivate the use of a time-dependent baseline intensity in our specification in Equation 4.4, we examined the average hourly intensity of our measured empirical event processes. Figure 8 shows the average intensity of all events for a candidate stock (GRT) for each hour of the trading day, averaged over all days in the respective datasets. Both datasets indicate a distinct $U$-shape for average intensities, which a constant baseline intensity will fail to capture.

Figure 9 illustrates the effect of different kernels when assuming a 3-period piecewise linear baseline intensity. We see that a simple *morning*, *noon* and *afternoon* distinction yields calibrated intensities which match the expected $U$-shape exhibited in the empirical data in Figure 8. The shapes of these curves are similar across all three kernels, although we note that as the number of exponentials is increased, we see a decline in the both the mean and variation of the exogeneity. This suggests that the higher number of exponentials permits more explanatory power for cross-exciting effects of events, rather than absorbing these effects into the baseline intensity. Combined with the *goodness-of-fit* tests in Section 6, this confirms that the drivers of observed event intensities in our data appear to have a higher attribution to self- and cross-exciting effects, as captured by the 3-exponential kernel, than that suggested by fewer exponentials.

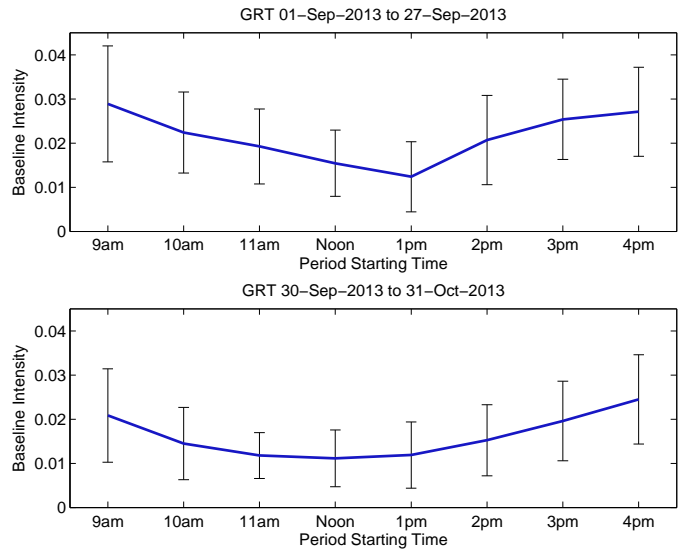

Fig. 8. **Average hourly baseline intensity for all events** Blue line indicates average for a given hour, with the error bars reflecting the variation over the days in the sample.

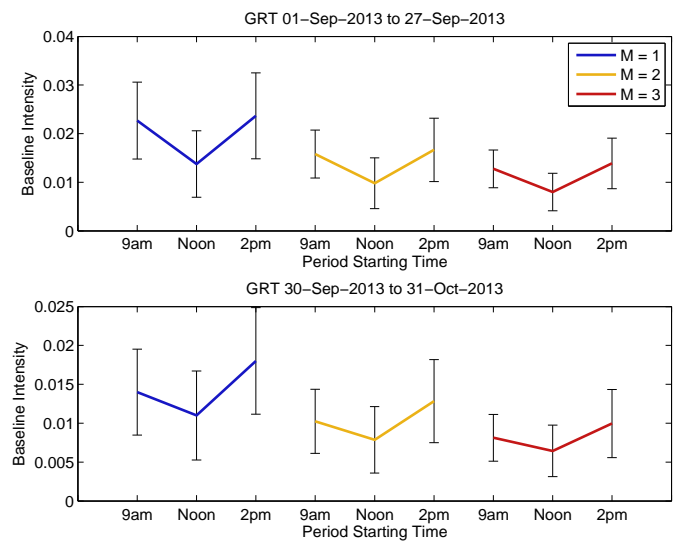

Fig. 9. **Baseline intensity by kernel**. The trading day is divided into 3 periods (*morning*, *noon*, *afternoon*), with piecewise linear intensity. Coloured lines correspond to mean exogenous intensities for the given periods, calibrated for each kernel, with the error bars reflecting daily variation.

18   *R. Martins, D. Hendricks*

**Branching Ratios**  Recalling the definition of the branching ratio given by Equation 4.1, we observe the self-exciting branching ratios calculated from the calibrated parameters in Figure 10, with their summary statistics presented in Tables 7 and 8. Intuitively, the branching ratios are higher for the bids and offers, for which there are far more events per day, in either dataset, than there are trades. As with the exogenous intensities, we see that the second dataset has lower branching ratios, corresponding to the lower event rates. Curiously, we observe that in contrast to the exogenous intensities that decline with the number of exponentials, the branching ratios actually increase as we increase the number of exponentials; from this we conjecture that increasing the number of exponentials in the model allows us to capture more of the individual self- and mutual-excitation effects, leaving less of the effective intensity absorbed by the exogenous factor.

|   | **Event 1** |   | **Event 2** |   | **Event 3** |   | **Event 4** |   |
|---|---|---|---|---|---|---|---|---|
| **M** | **Mean** | **Std** | **Mean** | **Std** | **Mean** | **Std** | **Mean** | **Std** |
| **1** | 0.2158 | 0.0826 | 0.1703 | 0.0626 | 0.3990 | 0.0939 | 0.4377 | 0.1489 |
| **2** | 0.2911 | 0.0972 | 0.2430 | 0.0753 | 0.4179 | 0.1212 | 0.4756 | 0.1416 |
| **3** | 0.3055 | 0.0863 | 0.2801 | 0.0740 | 0.4320 | 0.1184 | 0.4938 | 0.1347 |

Table 7. **Branching Ratio Statistics, GRT 01 September 2013 to 27 September 2013.** Shows the mean and standard deviation (std) of the distribution of branching ratios for each event, computed for each day in the dataset.

|   | **Event 1** |   | **Event 2** |   | **Event 3** |   | **Event 4** |   |
|---|---|---|---|---|---|---|---|---|
| **M** | **Mean** | **Std** | **Mean** | **Std** | **Mean** | **Std** | **Mean** | **Std** |
| **1** | 0.1705 | 0.0985 | 0.2003 | 0.0744 | 0.3212 | 0.1186 | 0.3413 | 0.1223 |
| **2** | 0.2346 | 0.1135 | 0.1941 | 0.0720 | 0.3724 | 0.1130 | 0.3897 | 0.1048 |
| **3** | 0.2821 | 0.1080 | 0.2520 | 0.0965 | 0.3860 | 0.0996 | 0.4256 | 0.1102 |

Table 8. **Branching Ratio Statistics, GRT 30 September 2013 to 31 October 2013.** Shows the mean and standard deviation (std) of the distribution of branching ratios for each event, computed for each day in the dataset.

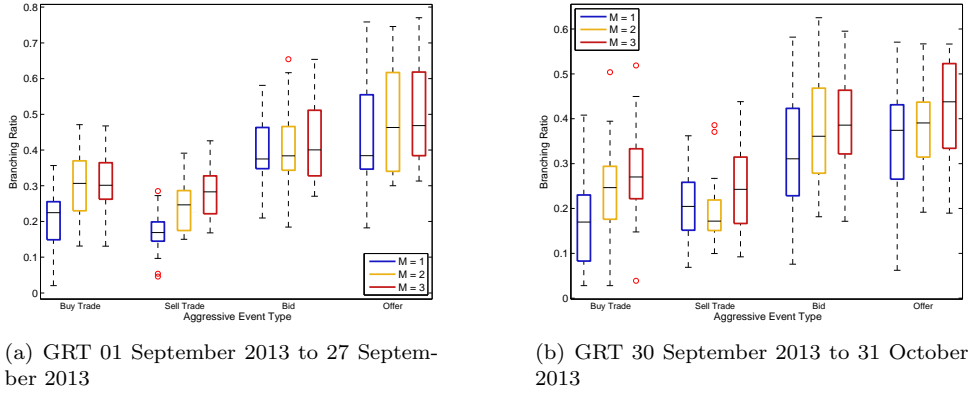

(a) GRT 01 September 2013 to 27 September 2013

(b) GRT 30 September 2013 to 31 October 2013

Fig. 10. **Box and whisker plot of branching ratios by kernel.** Horizontal black lines represent medians, box edges the quartiles, whiskers the furthest non-outlying data points, and red circles the outliers.

**Half-Lives** Recalling the definition of the half-life given by Equation 4.3, we observe the half-lives calculated from the calibrated parameters in Figure 11, with their summary statistics also presented in Tables 9 and 10. We note that the calculated half-lives appear to increase as the number of kernels is increased in each dataset. Given the superior fit of the 3-exponential kernel, demonstrated in Section 7.2.2, this suggests that the persistence of self- and mutual-excitation effects may be longer than is apparent under other model formulations. If we compare the two datasets in Tables 9 and 10, we note that there were significantly higher half-lives across all event types in the second period, with a higher variability. We note that on 30 September 2013, there was a significant change in the trading fee structure at the Johannesburg Stock Exchange, where a previous cost floor on exchange-related trading cost was replaced with a value-based cost, reducing the effective cost of small-value trades (JSE (2013a,b, 2014)). This appears to have significantly changed the activity of the various event types in the limit order book, especially the decay time of self- and mutual-excitation effects. Longer term studies should investigate the drivers of observed changes in limit order book dynamics from a resiliency perspective.

|   | Event 1 | | Event 2 | | Event 3 | | Event 4 | |
|---|---|---|---|---|---|---|---|---|
| **M** | **Mean** | **Std** | **Mean** | **Std** | **Mean** | **Std** | **Mean** | **Std** |
| **1** | 1.8904 | 4.9476 | 0.4765 | 1.5974 | 2.1506 | 2.1701 | 2.1929 | 3.7094 |
| **2** | 16.4678 | 24.3440 | 7.9023 | 9.5129 | 7.3177 | 5.7793 | 10.6777 | 10.1756 |
| **3** | 22.1760 | 27.4824 | 16.4143 | 14.9053 | 15.1500 | 10.0538 | 18.3472 | 11.1086 |

Table 9. **Half-life Statistics, GRT 01 September 2013 to 27 September 2013.** Shows the mean and standard deviation (std) of the distribution of half-lives (in seconds) for each event, computed for each day in the dataset.

|   | Event 1 | | Event 2 | | Event 3 | | Event 4 | |
|---|---|---|---|---|---|---|---|---|
| **M** | **Mean** | **Std** | **Mean** | **Std** | **Mean** | **Std** | **Mean** | **Std** |
| **1** | 0.5612 | 0.9662 | 1.3043 | 2.5518 | 1.2565 | 1.4263 | 1.9951 | 3.4089 |
| **2** | 38.1799 | 149.2049 | 2.4064 | 4.4341 | 13.3515 | 19.9383 | 13.5566 | 17.5231 |
| **3** | 56.0999 | 104.3957 | 43.6318 | 125.2765 | 28.2081 | 36.7412 | 35.8443 | 64.7057 |

Table 10. **Half-life Statistics, GRT 30 September 2013 to 31 October 2013.** Shows the mean and standard deviation (std) of the distribution of half-lives (in seconds) for each event, computed for each day in the dataset.

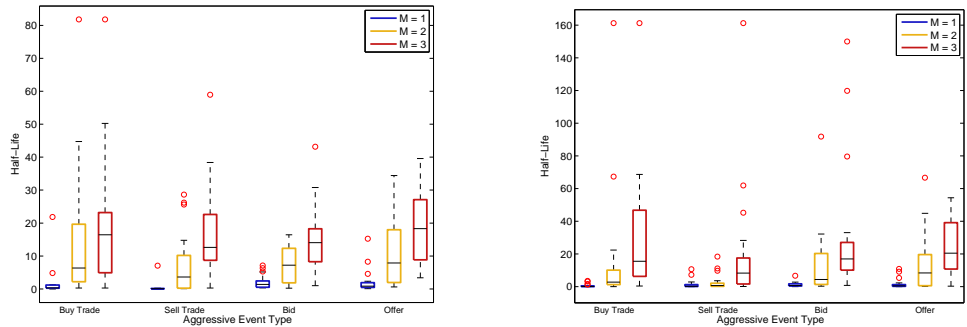

(a) GRT 01 September 2013 to 27 September 2013

(b) GRT 30 September 2013 to 31 October 2013

Fig. 11. **Box and whisker plot of half-lives by kernel.** Horizontal black lines represent medians, box edges the quartiles, whiskers the furthest non-outlying data points, and red circles the outliers.

## 8. Conclusion

The key investigation in this paper is to determine the statistical significance of multivariate Hawkes processes fitted to limit order book data obtained from the Johannesburg Stock Exchange (JSE). Given model specifications which use piecewise-linear exogeneity factors, and one- two- and three-exponential kernels to govern endogeneity of the intensity, we tested the residuals of calibrated models to assess their distribution and stationarity. We found that the sum-of-three-exponentials kernel offers the best fit to a four-variate Hawkes process, calibrated to key liquidity demand and replenishment events extracted from JSE limit order book data. This offers a more accurate quantification of order book resiliency using calibrated amplitude and time-scale parameters of liquidity replenishment intensities.

**Acknowledgments**

This work is based on the research supported in part by the National Research Foundation of South Africa (Grant number 89250). The conclusions herein are due to the authors and the NRF accepts no liability in this regard. The authors thank Diane Wilcox and Tim Gebbie for their comments and suggestions.

**References**

Abergel, F., Anane, M., Chakraborti, A., Jedidi, A., Toke, I., 2015. Limit order books. Working paper.
URL http://fiquant.mas.ecp.fr/wp-content/uploads/2015/10/Limit-Order-Book-modelling.pdf

Bacry, E., Dayri, K., Muzy, J., 2012. Non-parametric kernel estimation for symmetric hawkes processes. application to high frequency financial data. The European Physical Journal B **85** (157).

Bacry, E., Mastromatteo, I., Muzy, J., 2015. Hawkes processes in finance. Market Microstructure and Liquidity **1** (1).

Bacry, E., Muzy, J., 2014a. Hawkes model for price and trades high-frequency dynamics. Quantitative Finance **14** (7), 1147–1166.

Bacry, E., Muzy, J., 2014b. Second order statistics characterization of hawkes processes and non-parametric estimation. Working paper.
URL http://arxiv.org/pdf/1401.0903.pdf

Biais, B., Hillion, P., Spatt, C., 1995. An empirical analysis of the limit order book and the order flow in the paris bourse. Journal of Finance **50**, 1655–1689.

Bowsher, C., 2005. Modelling security market events in continuous time: Intensity-based, multivariate point process models. Journal of Econometrics **141** (2), 876–912.

Dassios, A., Zhao, H., 2013. Exact simulation of hawkes process with exponentially decaying intensity. Electronic Communications in Probability **18** (62), 1–13.

Degryse, H., deJong, F., Ravenswaaij, M., Wuyts, G., 2005. Aggressive orders and the resiliency of a limit order market. Review of Finance **9** (2), 201–242.

Engle, R., Russell, J., 1998. Autoregressive conditional duration: A new model for irregularly spaced transaction data. Econometrica **66**, 11271162.

Filimonov, V., Sornette, D., 2012. Quantifying reflexivity in financial markets: Toward a prediction of flash crashes. Physical Review E **85** (5), 1065–1073.

Filimonov, V., Sornette, D., 2013. Apparent criticality and calibration issues in the hawkes self-excited point process model: application to high-frequency financial data. Working paper.
URL http://arxiv.org/abs/1308.6756

Gould, M., Porter, M., Williams, S., McDonald, M., Fenn, D., Howison, S., 2013. Limit order books. Working paper.
URL http://arxiv.org/abs/1012.0349

Hardiman, S., Bercot, N., Bouchaud, J., 2013. Critical reflexivity in financial markets: a hawkes process analysis. The European Physica Journal B **86** (10), 1–9.

Hardiman, S., Bouchaud, J., 2014. Branching ratio approximation for the self-exciting hawkes process. Working paper.
URL http://arxiv.org/abs/1403.5227

Harvey, M., Hendricks, D., Gebbie, T., Wilcox, D., 2016. Deviations in expected price impact for small transaction volumes under fee restructuring. Working paper.
URL http://arxiv.org/abs/1602.04950

Hawkes, A., 1971. Spectra of some self-exciting and mutually-exciting point processes. Biometrika **58** (1), 83–90.

JSE, 2013a. Equity market price list (retrieved: 01/02/2016).
URL http://tinyurl.com/gqya25y

JSE, 2013b. Jse equity market transaction billing model methodology change notice, jse market notice no. 136 (retrieved: 01/02/2016).
URL http://tinyurl.com/hvht39j

JSE, 2014. Equity market price list 2014 v1.1 (retrieved: 01/02/2016).
URL http://tinyurl.com/jdlt38x

Kolmogorov, A., 1933. Sulla determinazione empirica di una legge di distribuzione. G. Ist. Ital. Attuari **4**, 83–91.

Kwiatowski, D., Phillips, P., Schmidt, P., Shin, Y., 1992. Testing the null hypothesis of stationarity against the alternative of a unit root. Journal of Econometrics 54 (1-3), 159–178.

Lallouache, M., Challet, D., 2016. The limits of statistical significance of hawkes processes fitted to financial data. Quantitative Finance **16** (1), 1–11.

Large, J., 2007. Measuring the resiliency of an electronic limit order book. Journal of Financial Markets **10**, 1–25.

Lewis, P., Shedler, G., 1979. Simulation of nonhomogeneous poisson processes by thinning. Naval Research Logistics Quarterly **26** (3), 403–413.

Ljung, G., Box, G., 1978. On a measure of a lack of fit in time series models. Biometrika **65** (2), 297–303.

Martins, R., 2015. The statistical significance of mutually-exciting hawkes processes fitted to jse data. AMF Honours Project, University of the Witwatersrand, Supervised by D. Hendricks.

Mazibuko, K., 2014. Quantifying resiliency of the jse limit order book following large trades. AMF Honours Project, University of the Witwatersrand, Supervised by D. Hendricks.

Ogata, Y., 1988. Statistical models for earthquake occurences and residual analysis for point processes. Journal of the American Statistical Association **83** (401), 9–27.

Ogata, Y., 1999. Seismicity analysis through point process modelling: a review. Pure and Applied Geophysics **155** (2), 471–507.

Ozaki, T., 1979. Maximum likelihood estimation of hawkes self-exciting point process. Annals of the Institute of Statistical Mathematics **31** (1), 145–155.

Smirnov, N., 1948. Table for estimating the goodness of fit of empirical distributions. Annals of Mathematical Statistics **19**, 279–281.

Vere-Jones, D., 1970. Stochastic models for earthquake occurence. Journal of the Royal Statistical Society, Series B (Methodological) **32** (1), 1–62.